# Low sintering temperature of Mg-Cu-Zn ferrites prepared by the citrate precursor method

Yomen Atassi*[1], and Mohammad Tally
Department of applied physics, Higher Institute for Applied Sciences and Technology, HIAST, P.O. Box 31983, Damascus, Syria.

Abstract:
Mg-Cu-Zn ferrite was prepared through a wet synthetic method by a self-combustion reaction directly from a citrate precursor. The as-synthesized powders were sintered at $750°C$ for only 2h. XRD patterns and FTIR spectra confirm the formation of single phase Mg-Cu-Zn ferrite after combustion. To the best of our knowledge, this is the first time that Mg-Cu-Zn ferrite is sintered at such low temperature. The sintering process increased the crystallinity of the solid and the domain sizes.

## 1. Introduction:

With the rapid development of mobile communication and information technology, the electronic devices with small size, low cost and high performance are quite demanded [1]. Recently, surface mounting devices (SMD) have been rapidly developed using multilayer chip inductors (MLCI). These chip inductors are produced by coating ferrite and electrical paste alternately and then co-firing. Since Ag is usually used as an electrode material, high temperature co-firing causes an unexpected lowering of the inductance due to the interfacial reaction (via diffusion) between the ferrite and the Ag electrode. This interfacial reaction can be suppressed [2] by co-firing at a temperature lower than the melting point of Ag (approximately $960°C$). Therefore, low-temperature sintering is of great importance to suppress the interfacial diffusion. Ni-Cu-Zn ferrite is usually used as a magnetic material for multilayer chip inductors due to its lower sintering temperature and better properties at high frequency than Ni-Zn ferrite. But it was found that these ferrites are comparatively sensitive to stress and magnetic properties easily changed or deteriorated by the stress caused at the internal electrode [2].

In order to overcome these problems, Mg-Cu-Zn ferrites were found to be suitable [3]. Normally, Mg-Cu-Zn ferrites were sintered at temperatures higher than $1000°C$. In order to use these ferrites in multilayer chip components, the sintering temperature must not be over the melting point of Ag.

The aim of this work is to present a novel and economical method of preparation of Mg-Cu-Zn ferrite by the citrate precursor method in order to achieve sintering at lower temperatures.

The citrate precursor method is a promising method in the synthesis of some technical ceramics [4,5]. Since all the reactants are solutions, they can be mixed at a level of molecule or atom, and the amount of the reactants can be controlled accurately. This wet chemical method has unique advantages over conventional sintering processes in terms of obtaining nanoparticles that can be densified easily at lower temperature.

---

* E-mail address: Yomen_Atassi@hiast.edu.sy

2. **Experimental procedure**:
**2.1. *Sample preparation*:**
$Mg_{0.25}Cu_{0.25}Zn_{0.50}Fe_2O_4$ was chosen to conduct this study. All used reagents were of AR grade. Aqueous solution of stoichiometric amounts of magnesium, copper, zinc and iron nitrates were reacted with citric acid in 1:1 molar ratio. pH of the solution was increased to 7 by addition of ammonium hydroxide to complete the reaction of the Mg-Cu-Zn citrate precursor.

The solution was evaporated very slowly over a period of ten to twelve hours to dryness. Viscosity and color changed as the sol turned into puffy, porous dry gel. As soon as the solvent removal is completed, dried precursor undergoes a self-ignition reaction to form a very fine powder known as as-synthesized powder.

The as-synthesized powder thus obtained was treated in a furnace at $250°C$ for 6 h followed by a further heat treatment at $380°C$ for 6 h to remove the residual carbon. Then the powder was divided into two patches, the first patch was pressed into pellets of 10 mm diameter and a thickness of 4 mm for dilatometer studies, and the second patch was pressed into pellets (of 25 mm at 200 MPa) and toroids (with internal and external diameters 10 and 25 mm, respectively, and a thickness of 4 mm) and sintered at $750°C$ for two hours in air.

**2.2. *Characterization*:**
The thermal decomposition of the as-synthesized powder was investigated by means of a differential thermal analyser (DTA, Linseis L62 thermal analyzer) at a heating rate of $10°C/min$ in air.

In order to determine the suitable sintering temperature, dilatometer studies were performed using (Setaram, TMA 92) at a heating rate of $10°C/min$ in nitrogen.

A computer interface X-ray powder diffractometer (Philips) with Cu $K\alpha$ radiation ($\lambda = 0.1542\,nm$) was used to identify the crystalline phase.

The data collection was over the 2-theta range of $20°$ to $80°$ in steps of $0.04°/sec$. Crystallite size was calculated from the width of the (311) line using the Scherrer formula corrected for instrumental broadening.

The IR spectra in the 400-2000 cm$^{-1}$ range were recorded at room temperature on the infrared spectrophotometer (Bruker, Vector22). For recording IR spectra, powders were mixed with KBr in the ratio 1:250 by weight to ensure uniform dispersion in the KBr pellet. The mixed powders were then pressed in a cylindrical die to obtain clean discs of approximately 1 mm thickness.

A definite number of turns were wound around each toroid to measure the inductance of the coil, *L*, in the range 10-2000 kHz.

The initial permeability $\mu_i$ was estimated from the relation:
$$\mu_i = \frac{D}{B\,n^2} 10^8 L \,,$$

where $D = \dfrac{a+b}{2}$ and $B = (b-a)h$ are the radius and average area of the toroidal samples, $a$ and $b$ being the internal and external radii of the toroid, $h$ the thickness of the toroid, and $n$ the number of turns.

## 3. Results and discussions:
### 3.1. *Thermal behavior*:

Figure (1) shows the DTA curve of the as-synthesized powder. There is a large exothermic peak in the range 320-380 °C which is assigned to the decomposition of the citrate complex.
Whereas the gentle, faint endothermic peak at 590°C corresponds to solid state reaction of the resulting oxides.

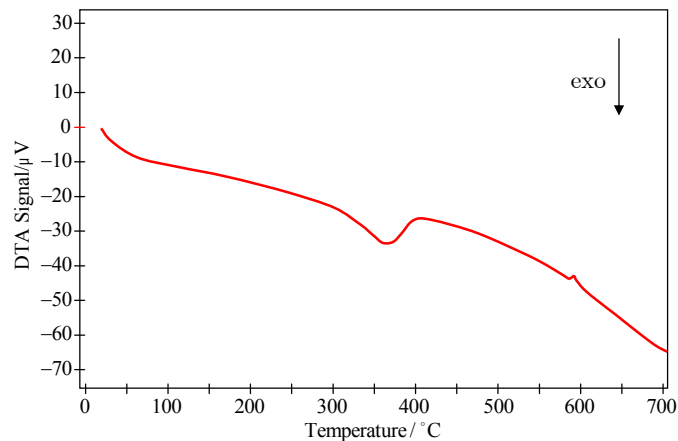

Figure (1): DTA curve of the as-synthesized Powder.

In order to determine the suitable temperature of densification, dilatometer curve of the pellets prepared from the powder treated at 380 °C was registered, Figure (2). It showed that the suitable sintering temperature is about 750°C.

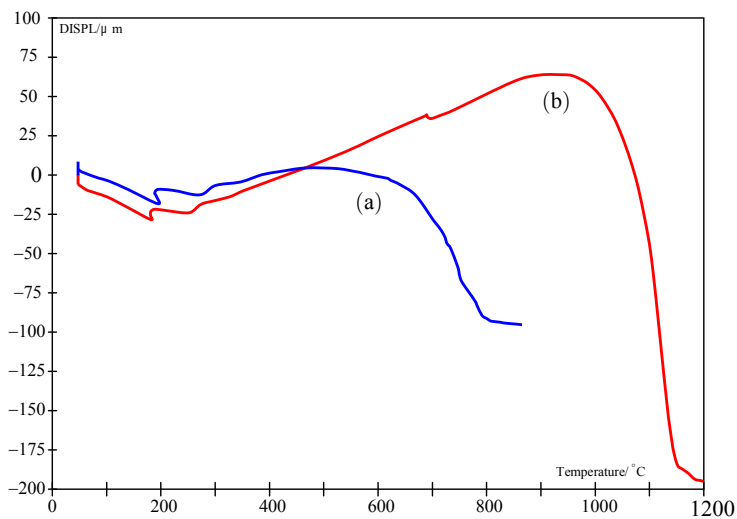

Figure(2): Dilatometer Diagram of:
(a) pellet of 10 mm diameter and a thickness of 4 mm treated at 250°C and 380°C.
(b) pellet of 10 mm diameter and a thickness of 6 mm previously treated by microwave till the disappearance of glow.

However, we noticed that the sintering temperature augmented to about $1200°C$ when we replaced the heat treatment of the as-synthesized powder by a microwave treatment till the disappearance of glow, in order to realize both the decomposition of the citrate and the removal of the residual carbon. We attribute this augmentation in the sintering temperature to the large size of grains due to microwave treatment [6].

### 3.2. X-ray patterns:
The X-ray diffraction patterns of the as-synthesized powder, the heat-treated powder and the sintered ceramics at $750°C$ for 2h. are shown in Figure (3). It is evident that the sintered sample contains only the spinel cubic ferrite. Mg-Cu-Zn ferrite is formed and all the peaks in the pattern match well with JCPDS card.

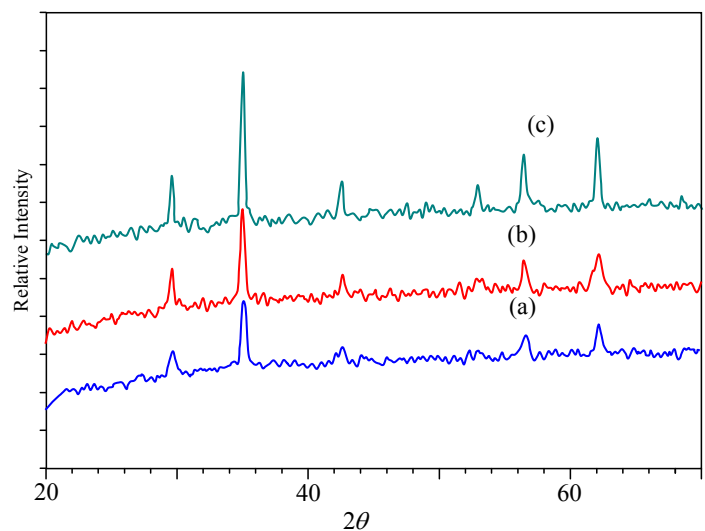

Figure (3): The X-ray diffraction patterns of: (a) the as-synthesized powder, (b) the heat-treated powder and (c) the sintered ceramics at $750°C$ for 2h.

There are no additional peaks corresponding to extra phases such as $\alpha\text{-}Fe_2O_3$ which often appear at low temperatures when the reaction is not complete. It may be mentioned that, synthesis of single phase Mg-Cu-Zn ferrite by conventional ceramic methods necessitate repeated grinding and calcinations at temperatures above $1000°C$ [7].
The sintering process increased the crystallinity of the solid and the domain sizes. Crystallite size was estimated at 24 nm.

### 3.2. IR Spectra:
In order to confirm the formation of the spinel phase and to understand the nature of the residual carbon in the samples, the FTIR spectra of the as-synthesized powders and thermally treated powder were recorded, Figure (4).
The as-synthesized sample shows characteristic absorptions of ferrite phase with a strong absorption around 600 cm$^{-1}$ and weak absorption in the range 410-450 cm$^{-1}$[8]. This difference in the band position is expected because of the difference in the $M^{n+}\text{-}O^{2-}$ distance for the octahedral and tetrahedral compounds. Waldron [9] studied the vibrational spectra of ferrites and attributed the sharp absorption band around 600 cm$^{-1}$ to the intrinsic vibrations of the tetrahedral groups and the other band the octahedral groups.

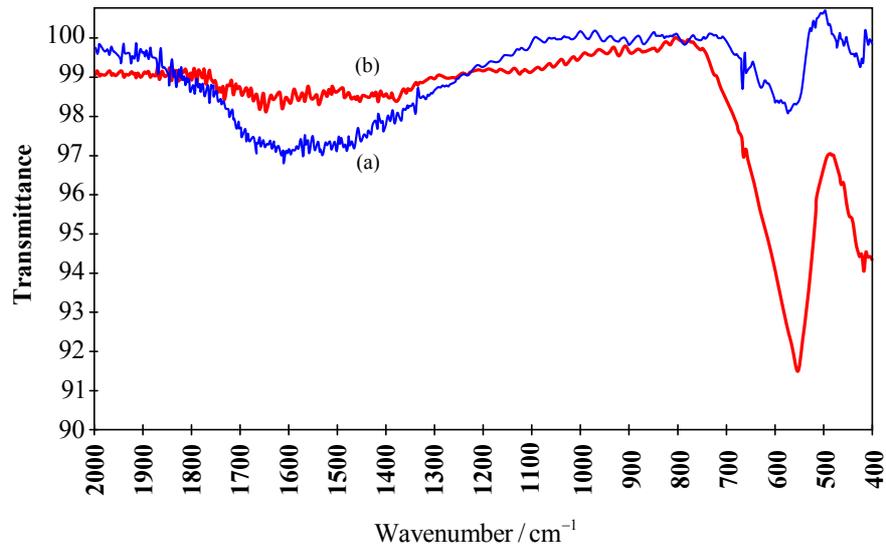

Figure (4): The FTIR spectra of: (a) the as-synthesized powder, (b) the heat-treated powder.

There are two weak and broad absorptions around 1400 and 1600 cm$^{-1}$ corresponding to the presence of small amount of residual carbon in the samples. These absorptions in the present case are very weak which indicates that the residual carbon has mostly burnt away during the self-combustion process. The wave numbers of these absorptions indicate that this carbon is in the form of complex carbonates. Heat treatment at $250°C$ for 6 h followed by a further heat treatment at $380°C$ for 6 h led to the disappearance of these absorptions indicating complete removal of residual carbon from the sample [10].

### 3.4. *Initial permeability*:
It's known that the permeability of polycrystalline ferrite can be described as the superposition of domain wall and pin rotation components. At small grain sizes, the grain become monodomain and the reversal of magnetization can only occur through spin rotation.

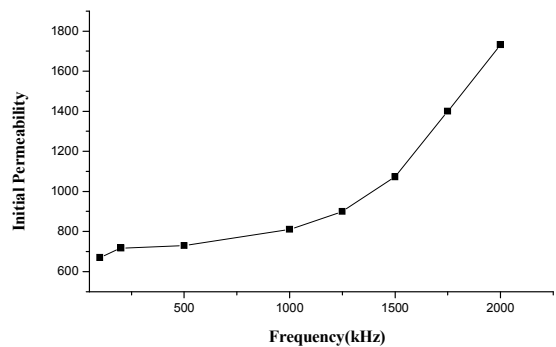

Figure (5): Initial permeability versus frequency.

Figure (5) shows the variation of the initial permeability with frequency in the region 10-2000 kHz. We connect the augmentation of the initial permeability near 2 MHz to the relaxation process in the RF region [11].

## 4. Conclusion:

The powder synthesized by the citrate precursor method demonstrated considerable sinterability. It's regarded that the fine particle morphology of the powder synthesized by this method is responsible for its higher sintering activity. The very fact that single phase ferrite could be obtained directly by citrate precursor without any additional heat-treatments above $750°C$ is a significant achievement considering the variety of applications of the Mg-Cu-Zn ferrite. The highly active powders could be sintered at relatively low temperatures to obtain highly dense and homogeneous polycrystalline ferrites for high frequency applications.